\documentclass[pra,reprint,amsmath,amssymb,nofootinbib]{revtex4-2}
\usepackage{graphics}
\usepackage{dsfont}
\usepackage{natbib}
\usepackage{hyperref}
\usepackage[mathscr]{eucal}
\usepackage{mathtools}
\usepackage{bm}
		
\begin{document}

\title{Measuring trajectories of environmental noise}

\author{Piotr Sza\'{n}kowski}\email{piotr.szankowski@ifpan.edu.pl}
\affiliation{Institute of Physics, Polish Academy of Sciences, al.~Lotnik{\'o}w 32/46, PL 02-668 Warsaw, Poland}

\begin{abstract}
In classical mechanics, a natural way to simplify a many-body problem is to ``replace'' some of the elements of the composite system with surrogate \textit{force fields}. In the realm of quantum mechanics, however, such a description is rarely compatible with the formalism of the theory. Nevertheless, the quantum version of external field models---the so-called \textit{noise representations}---can be employed in certain circumstances. The mathematics behind these models indicate that the appearing fields typically exhibit random fluctuations, hence, the name \textit{noise field} is more apt. In principle, measuring a classical force field is a trivial task; all that is needed is a probe equipped with an accelerometer that can be moved around while taking the measure of forces that affect it. Unfortunately, an analogous method cannot work in the quantum case. As indicated by the theory, the result of any measurement performed on the quantum system affected by a noise field always appears as an average over fluctuations, therefore, it is impossible to observe the noise fluctuations by measuring the state of the probe. Here, we demonstrate that this limitation can be circumvented and that it is possible to expose the noise field fluctuations to direct observation. This allows one to sample the noise trajectories (stochastic realizations), store them, and later use them to simulate the dynamics of open quantum systems affected by the noise.
\end{abstract}
		
\maketitle
	
\section{Introduction}

In recent years, we are witnessing a rapid pace of advancement in the field of quantum technologies. This progress is spearheaded by a considerable effort towards the design of nanoscale devices that employ genuine quantum effects to perform certain tasks with efficiency that, at least theoretically, surpasses the limits imposed by classical physics \cite{Acin18}; some examples include computing \cite{Preskill18}, metrology \cite{Giovannetti11, Pezze18,Demkowicz20}, and communication \cite{Gisin07, Kimble08, Northup14}. Arguably, the most significant obstacle standing between the theoretical concept of a quantum device and its practical implementation is the phenomenon of decoherence that results in the erosion of subtle, yet indispensable, quantum features. This hindrance is unavoidable in virtually any realistic setting because of the inevitable influence of the environment \cite{Schlosshauer07}. To understand, predict, and hopefully counteract the destructive effects of decoherence, a systematic description of the dynamics of a system open to the environment is required~\cite{Breuer02}. An alluring branch of the theory of open systems is the noise representation of the system--environment interaction, where the environment is treated as a driving force and the system's role is reduced to a passive recipient of this driving. Formally, such representation is implemented by replacing the full system--environment Hamiltonian with the system-only operator where the environmental degrees of freedom are ``replaced'' by an external noise---the so-called surrogate field\footnote{
Note that, traditionally, the surrogate field is referred to in the literature as the classical noise. Here, we adopt the stance advocated for in Ref.~\cite{Szankowski20} to phase out such a nomenclature. A number of examples shown in Ref.~\cite{Szankowski20} demonstrate that the use of the descriptor ``classical'' in this context is often inadequate and, in some cases, can even be misleading; e.g., the observation that the existence of the noise representation does not preclude the formation of system--environment entanglement, is a particularly compelling argument.
} $\xi$,
\begin{eqnarray}
\nonumber
&&\hat H_{SE} = \hat H_S\otimes\hat{\mathds{1}} + \hat V_S\otimes\hat V_E + \hat{\mathds{1}}\otimes\hat H_E \\
\label{eq:H_SE}
&&\phantom{H_{SE} = }
	\xrightarrow{ \substack{ \text{noise}\\ \text{representation}} }\hat H_t[\xi] = \hat H_S + \xi(t)\hat V_S.
\end{eqnarray}
Here, $\hat H_{S(E)}$ are the free Hamiltonians of system (environment) and $\hat V_{S(E)}$ are the system (environment) sides of the coupling. When noise representation is valid, the system dynamics can be simulated with the averaged unitary evolution driven by the surrogate field $\xi$,
\begin{eqnarray}
\nonumber
    \hat\rho_S(t) &=& \mathrm{tr}_E\left(e^{-it\hat H_{SE}}\hat\rho_S\otimes\hat\rho_E\, e^{it\hat H_{SE}}\right)\\
\label{eq:rhoS}
    &=&\overline{\hat U_t[\xi]\hat\rho_S\hat U^\dagger_t[\xi]}.
\end{eqnarray}
Here, $\mathrm{tr}_E(\ldots)$ is the partial trace over the $E$ subspace, $\overline{(\ldots)}$ is the average over realizations (trajectories) of the stochastic process $\xi$, and each individual realization is governed by simple (relative to the full $SE$ problem) unitary evolution operator conditioned by the trajectory~$\xi(\tau)$,
\begin{equation}
    \hat U_t[\xi] = \mathcal{T}e^{-i\int_0^t\hat H_\tau[\xi]d\tau},
\end{equation}
where $\mathcal{T}$ indicates the time-ordering operation. Recent developments \cite{Szankowski20} have provided an explanation of the physical origins of the noise representation and have quantified the sufficient conditions for the environment to facilitate the surrogate field that represents it. These explanations have been built around formal analogy between stochastic average and the partial trace operation. On the one hand, the average can be expressed in the terms of functional integral
\begin{equation}\label{eq:path_int}
    \overline{\hat U_t[\xi]\hat\rho_S\hat U^\dagger_t[\xi]}
    = \int\mathcal{D}\xi\, \mathcal{P}[\xi] \hat U_t[\xi]\hat\rho_S\hat U^\dagger_t[\xi],
\end{equation}
where the non-negative functional $\mathcal{P}[\xi]$ is the probability distribution for trajectories $\xi(t)$ (real-valued functions of time) of stochastic process $\xi$. On the other hand, as demonstrated in~\cite{Szankowski20}, it is always possible to rewrite the partial trace as
\begin{eqnarray}
\nonumber
&&\mathrm{tr}_E\left(e^{-it\hat H_{SE}}\hat\rho_S\otimes\hat\rho_Ee^{it\hat H_{SE}}\right)\\
\label{eq:trace}
&&\phantom{\mathrm{tr}_E(}   
	=\int\mathcal{D}\xi\int\mathcal{D}\zeta\,\mathcal{Q}[\xi,\zeta]
    \hat U_t[\xi]\hat\rho_S\hat U_t^\dagger[\zeta],
\end{eqnarray}
where the functional $\mathcal{Q}[\xi,\zeta]$ is complex-valued and, hence, it can be classified as a \textit{quasi-probability} distribution for the real-valued trajectories of the two-component ``quantum process'' $(\xi(\tau),\zeta(\tau))$. By comparing~\eqref{eq:path_int} and~\eqref{eq:trace}, one concludes that the noise representation is effected when the quasi-probability can be treated as a proper probability distribution, i.e., when
\begin{equation}\label{eq:validity_func}
\mathcal{Q}[\xi,\zeta] = \delta(\zeta-\xi)\mathcal{P}[\xi] \geqslant 0.
\end{equation}

Note that $\mathcal{Q}$ is not directly related to the influence functional used in the description of open-system dynamics within Feynman's path integral formalism~\cite{Vernon63,Caldeira83,Legget87}. In more technical terms, the paths in Feynman's formalism and the processes $\xi,\zeta$ in~\eqref{eq:trace} are fundamentally different entities: the former refer to dynamical variables associated with the Lagrangians assigned to quantum systems $S$ and $E$, while the latter can be interpreted as trajectories traced through the domain of the eigenvalues of the $E$ side of the coupling $\hat V_E$. More importantly, the influence functional contains the state $\hat\rho_E$, the Hamiltonian $\hat H_E$, and the $SE$ interaction (thus, making it dependent on $S$), while $\mathcal{Q}$ leaves out the interaction term (it is incorporated into $\hat U_t$, instead); this is a crucial difference.

The key feature of the formulation~\eqref{eq:trace} is that the properties of quasi-probability $\mathcal{Q}$ are determined exclusively by the environment, i.e., $\mathcal{Q}$ is independent of the open system $S$~\cite{Szankowski20}. Therefore, the validity of the noise representation (i.e., whether $\mathcal{Q}$ is a proper probability distribution) is a feature of the environment, and the form of the resultant surrogate field is established only by the dynamical laws governing the environmental degrees of freedom (i.e., the triplet of operators $\hat H_E$, $\hat\rho_E$, and $\hat V_E$). It also means that the validity of noise representation is \textit{not} contingent on the effects of interaction between $E$ and any particular $S$, i.e., it does not matter whether $S$ exerts back-action onto $E$, or if $S$ entangles with $E$, etc.~\cite{Szankowski20}. Consequently, when the noise representation exists for a given environment, it is an \textit{objective} representation---the dynamics of any system $S$ defined by the choice of $\hat H_S$, $\hat\rho_S$, and $\hat V_S$, that is coupled to $E$ via a given environment-side operator $\hat V_E$, is simulable with the same surrogate field that effectively replaces $\hat V_E$. This is the main reason why noise representations are so useful whenever they can be implemented: once one determines the objective surrogate field generated by the environment, describing the dynamics of any system open to this environment is significantly simplified.

From the point of view of practical implementation, the main challenge in executing the noise representation simulation is the average operation. The functional integral form~\eqref{eq:path_int}, as well as the probability distribution functional $\mathcal{P}[\xi]$, are merely formal theoretical tools that are never suitable for actual calculations. The solution that is practical exploits the equivalence between the average over the probability distribution and the average over the ensemble of samples drawn from this distribution,
\begin{eqnarray}
\nonumber
    \overline{\hat U_t[\xi]\hat\rho_S\hat U^\dagger_t[\xi]}
    &=&\int\mathcal{D}\xi\,\mathcal{P}[\xi]\hat U_t[\xi]\hat\rho_S\hat U^\dagger_t[\xi]\\
    &=& \lim_{N\to\infty}\frac{1}{N}\sum_{j=1}^N \hat U_t[\xi^{(j)}]\hat\rho_S\hat U^\dagger_t[\xi^{(j)}],
\end{eqnarray}
where the trajectories $\xi^{(j)}(\tau)$ belong to the set $\mathcal{E}_N = \{\xi^{(j)}(\tau)\}_{j=1}^N$---the ensemble of sample trajectories of stochastic process $\xi$ drawn at random from the distribution $\mathcal{P}[\xi]$. Of course, assembling the infinite number of samples is an impossibility, and so, in practice, one resorts to the approximation
\begin{equation}
\overline{\hat U_t[\xi]\hat\rho_S\hat U_t^\dagger[\xi]}\approx \frac{1}{N_e}\sum_{j=1}^{N_e} \hat U_t[\xi^{(j)}]\hat\rho_S\hat U^\dagger_t[\xi^{(j)}],
\end{equation}
where $N_e$ is the number of samples in the ensemble $\mathcal{E}_{N_e}$ of finite size; the larger the number of samples, the more accurate the approximation. Therefore, to find the average, instead of performing the abstract functional integral, one takes the ensemble of sample trajectories $\mathcal{E}_{N_e}$ and calculates the unitary evolution $\hat\rho_S^{(j)}(t) = \hat U_t[\xi^{(j)}]\hat\rho_S\hat U^\dagger_t[\xi^{(j)}]$ for every trajectory $\xi^{(j)}(\tau)$ (e.g., by numerically integrating the corresponding von Neumann equation). The results are then stored in the ensemble of density matrices $\{\hat\rho_S^{(j)}(t)\}_{j=1}^{N_e}$. The individual members of this ensemble are meaningless and their only purpose is to calculate the physically meaningful average that approximates the system state $(1/N_e)\sum_{j=1}^{N_e}\hat\rho_S^{(j)}(t) \approx \hat\rho_S(t)$.

Clearly, for this procedure to work, first, one needs to populate the ensemble of trajectories. According to Ref.~\cite{Szankowski20}, in principle, the trajectories can be drawn from so-called \textit{joint probability distribution} $p_E^{(k)}(\xi_{k}t_k;\ldots;\xi_{2}t_2;\xi_1t_1)$ that describes the probability for the surrogate field $\xi$ to have value $\xi_1$ at the initial time $t_1$, followed by $\xi_{2}$ at $t_2 > t_1$, followed by $\xi_3$ at $t_3 > t_2$, etc. The joint probability distributions form an infinite family $\{p_E^{(k)}\}_{k=1}^\infty$ that fully defines the stochastic process $\xi$, and it can be used as an alternative to the description in terms of the distribution functional $\mathcal{P}[\xi]$. The fundamental advantage one gets for using such a mode of description is that, unlike the abstract functional, joint distributions are standard functions given by closed analytical formulas~\cite{Szankowski20},
\begin{eqnarray}
\nonumber
&&p^{(k)}_E(\xi_{k}t_k;\ldots;\xi_1t_1) = \sum_{\substack{n_1:\\\xi_1 =v_{n_1}}}\cdots\sum_{\substack{n_k:\\\xi_k =v_{n_k}}}\langle n_1|\hat\rho_E(t_1)|n_1\rangle\\
\label{eq:joint}
&&\phantom{p^{(k)}_E(\xi_{k}t_k;\ldots;\xi_1)}
	\times\prod_{l=1}^{k-1}|\langle n_{l+1}|e^{-i(t_{l+1}-t_{l})\hat H_E}|n_{l}\rangle|^2,
\end{eqnarray}
where $\hat\rho_E(t) = e^{-it\hat H_E}\hat\rho_E\,e^{it\hat H_E}$ and $\{v_n\}_n$, $\{|n\rangle\}_n$ are (possible degenerate) eigenvalues and eigenstates of the coupling operator $\hat V_E$ that is to be replaced in the simulation by the surrogate field $\xi$. However, calculating any one of those distributions is almost as difficult as solving the original system--environment problem (one needs to know $\hat H_E$, $\hat V_E$, and $\hat\rho_E$, and be able to diagonalize the Hamiltonian). This is not entirely surprising; after all, the surrogate field, when it exists, is an exact representation of the environmental degrees of freedom. This difficulty could be circumvented if one was able to draw the sample trajectories from some other, ideally more accessible, source. In Sec.~\ref{sec:proof}, we demonstrate that the sequential measurements of the environmental observable $\hat V_E$ can be used as such a source of the surrogate field trajectories. Therefore, for the environments that facilitate noise representation, one can setup the ensemble of trajectories---and thus enable future simulations of open-system dynamics---by directly measuring the sample trajectories instead of drawing them from a calculated probability distribution.

\section{Trajectory sampling: the proof of concept}\label{sec:proof}

Here, we demonstrate that, assuming a valid noise representation, the result of a sequence of \textit{projective measurements}~\cite{Peres06} of observable $\hat V_E$ performed directly on the environment $E$, can be considered as a sample trajectory of the surrogate field $\xi$. To achieve this goal, we make extensive use of some technical results obtained previously in Ref.~\cite{Szankowski20}, that we cite without redundant derivation. 

First, we define the spectral decomposition of the observable of interest,
\begin{eqnarray}
\nonumber
\hat V_E &=& \sum_n v_n |n\rangle\langle n|\\
\label{eq:spectral}
&=& \sum_{\xi\in\Omega}\xi\sum_{n:v_n=\xi}|n\rangle\langle n| \equiv \sum_{\xi\in\Omega}\xi\, \hat \Pi(\xi),
\end{eqnarray}
where $\Omega$ is the set of all its unique eigenvalues (the spectrum). The operator $\hat V_E$ above is the same as the environmental side of the coupling in Eq.~\eqref{eq:H_SE} that is effectively superseded by the surrogate field $\xi$ when the noise representation is valid. 

Next, in order to showcase how exactly the assumption of a valid noise representation comes into play, we recall the \textit{joint quasi-probability} decomposition of the state of arbitrary system $S$ open to environment $E$. Following the results of Ref.~\cite{Szankowski20}, the density matrix of $S$ in the interaction picture, $\hat\rho_S^I(t) = e^{it\hat H_S}\hat\rho_S(t)e^{-it\hat H_S}$, can be written as
\begin{eqnarray}
\nonumber
&&\hat\rho^I_S(t) = e^{it\hat H_S}\mathrm{tr}_E\left(e^{-it\hat H_{SE}}\hat\rho_S\otimes\hat\rho_E\,e^{it\hat H_{SE}}\right)e^{-it\hat H_S}\\
\nonumber
&&=\sum_{k=0}^\infty (-i)^k \int_0^t\!\!dt_k \int_0^{t_k}\!\!dt_{k-1}\cdots\int_0^{t_{2}}\!\!dt_1
	\!\sum_{\xi_1,\zeta_1\in\Omega}\!\cdots\!\sum_{\xi_k,\zeta_k\in\Omega}\\
\label{eq:quasi-prob_decomp}
&&\phantom{\sum}\times
	q_E^{(k)}(\xi_k\zeta_kt_k;\ldots;\xi_1\zeta_1 t_1)
	\left(\prod_{l=k}^1\mathscr{W}(\xi_l\zeta_lt_l)\right)\hat\rho_S,
\end{eqnarray}
where the super operators acting on the initial state are given by the $S$-side of the $SE$ coupling, $\mathscr{W}(\xi\zeta t)\hat A = \xi \hat V_S(t)\hat A - \zeta \hat A\hat V_S(t)$ with $\hat V_S(t) = e^{it\hat H_S}\hat V_S e^{-it\hat H_S}$, and the symbol $\prod_{l=k}^1\mathscr{A}_l$ indicates the product where super-operators are arranged in the decreasing order of the index, $\mathscr{A}_k\mathscr{A}_{k-1}\cdots\mathscr{A}_1$. The members of the infinite family of nonpositively defined functions $\{ q_E^{(k)}\}_{k=1}^\infty$---the \textit{joint quasi-probability distributions}---are determined exclusively by the environment and their explicit form reads
\begin{widetext}
\begin{eqnarray}
\nonumber
q_E^{(k)}(\xi_k\zeta_kt_k;\ldots;\xi_1\zeta_1t_1)&=&
	\mathrm{tr}\left[
		\left(\prod_{l=k}^1 e^{it_l\hat H_E}\hat\Pi(\xi_l)e^{-it_l\hat H_E}\right)\hat\rho_E\left(\prod_{l'=1}^k e^{it_{l'}\hat H_E}\hat\Pi(\zeta_{l'})e^{-it_{l'}\hat H_E}\right)
	\right]\\
\label{eq:q_E}
&=&\delta_{\xi_k,\zeta_k}\mathrm{tr}\left[\hat\Pi(\xi_k)
		\left(\prod_{l=k-1}^1 e^{i(t_{l}-t_{l+1})\hat H_E}\hat\Pi(\xi_l)\right)
		\hat\rho_E(t_1)\left(\prod_{l'=1}^{k-1}\hat\Pi(\zeta_{l'})e^{i(t_{l'+1}-t_{l'})\hat H_E}\right)\right],
\end{eqnarray}
\end{widetext}
where the symbol $\prod_{l=k}^1\hat A_l$ ($\prod_{l'=1}^k\hat A_{l'}$) indicates the product where operators are arranged in the decreasing (increasing) order of the index $\hat A_k\hat A_{k-1}\cdots\hat A_1$ ($\hat A_1\cdots\hat A_k$).

The joint quasi-probability decomposition~\eqref{eq:quasi-prob_decomp} is a practical alternative to the functional-integral description~\eqref{eq:trace}; essentially, $\{q_E^{(k)}\}_{k=1}^\infty$ is to $\mathcal{Q}[\xi,\zeta]$ as $\{p_E^{(k)}\}_{k=1}^\infty$ is to $\mathcal{P}[\xi]$ [see Eq.~\eqref{eq:joint}]. When translated into the language of joint distributions, the sufficient condition for valid noise representation $\mathcal{Q}[\xi,\zeta] \approx \delta(\zeta-\xi)\mathcal{P}[\xi]$ becomes
\begin{eqnarray}
\nonumber
&&q_E^{(k)}(\xi_k\zeta_kt_k;\ldots;\xi_1\zeta_1t_1)\approx \left(\prod_{l=1}^k\delta_{\xi_l,\zeta_l}\!\right)\!p_E^{(k)}(\xi_kt_k;\ldots;\xi_1t_1),\\[-0.1cm]
\label{eq:validity}
\end{eqnarray}
for all $k=1,2,\ldots$. When the above condition is satisfied, then 
\begin{eqnarray}
\nonumber
\hat\rho^I_S(t) &=& \sum_{k=0}^\infty (-i)^k \int_0^tdt_k\cdots\int_0^{t_2}dt_1\sum_{\xi_1\ldots\xi_k\in\Omega}\\
\nonumber
&&\phantom{\sum}\times p_E^{(k)}(\xi_kt_k;\ldots;\xi_1t_1)\left(\prod_{l=k}^1\mathscr{W}(\xi_l\xi_lt_l)\right)\hat\rho_S\\
\nonumber
&=&\sum_{k=0}^\infty (-i)^k \int_0^tdt_k\cdots\int_0^{t_2}dt_1 \,\overline{\xi(t_k)\cdots\xi(t_1)}\\
\nonumber
&&\phantom{\sum}
	\times [\hat V_S(t_k),\cdots [\hat V_S(t_2), [\hat V_S(t_1),\hat\rho_S ]\;]\cdots\;]\\[.1cm]
&=&e^{i t \hat H_S}\overline{\hat U_t[\xi]\hat\rho_S \hat U_t^\dagger[\xi]}e^{-it\hat H_S},
\end{eqnarray}
where we have used the relation $\mathscr{W}(\xi_l\xi_l t_l)\hat A = \xi_l [\hat V_S(t_l),\hat A]$, and the \textit{moment} of stochastic process $\xi$ reads
\begin{eqnarray}
\nonumber
    \overline{\xi(t_k)\cdots\xi(t_1)} &=&\sum_{\xi_1,\ldots,\xi_k\in\Omega}\left(\prod_{l=1}^k\xi_l\right) p_E^{(k)}(\xi_kt_k;\ldots;\xi_1t_1),\\[-.1cm]
\label{eq:moment}
\end{eqnarray}
hence the interpretation that $p_E^{(k)}$ is the joint distribution that describes the probability of the trajectory passing through value $\xi_1$ at $t_1$ followed by $\xi_2$ at $t_2$, etc.

Of course, whether the relation~\eqref{eq:validity} holds depends only on the properties of~$E$ (hence, the objectivity of the noise representation), and thus, this equality can be invoked in any context---its utility is not limited to the quasi-probability decomposition of density matrices. We use this to our advantage as we now switch to a different setup where, instead of open-system dynamics, we examine the sequence of measurements performed directly on $E$.

The environment is initialized in the state $\hat\rho_E$ and its free evolution is governed by the Hamiltonian $\hat H_E$. Assume we can perform an ideal \textit{projective measurement} of physical quantity described by the observable $\hat V_E$ characterized by the spectral decomposition~\eqref{eq:spectral}. Then, according to the \textit{Born rule}, the probability of obtaining a particular eigenvalue $\xi_{t_1}$ as a result of the measurement carried out after duration $t_1$ has passed is given by $P(\xi_{t_1}) = \mathrm{tr}[\hat\Pi(\xi_{t_1})\hat\rho_E(t_1)]$, which, when compared with Eq.~\eqref{eq:q_E}, equals
\begin{equation}
P(\xi_{t_1}) =q^{(1)}_E(\xi_{t_1}\xi_{t_1}t_1).
\end{equation}
In accordance with the \textit{collapse postulate}, assuming that the duration of the measurement process is effectively instantaneous in comparison with the time scale set by $\hat H_E$, the \textit{a~posteriori} state of the environment continues to evolve as
\begin{eqnarray}
\nonumber
\hat\rho(t|\xi_{t_1}) &=&\frac{ e^{-i(t-t_1)\hat H_E}\hat\Pi(\xi_{t_1})\hat\rho_E(t_1)\hat \Pi(\xi_{t_1})e^{i(t-t_1)\hat H_E}}{q_E^{(1)}(\xi_{t_1}\xi_{t_1}t_1)}.\\[-.1cm]
\end{eqnarray}
When the projective measurement is performed again on the same system after certain period, then the (now conditional) probability of obtaining result $\xi_{t_2}$ at time $t_2>t_1$ (provided that $\xi_{t_1}$ was measured previously) is given by
\begin{widetext}
\begin{eqnarray}
P(\xi_{t_2}|\xi_{t_1})\! &=& \mathrm{tr}\!\left[\hat\Pi(\xi_{t_2})\hat\rho(t_{2}|\xi_{t_1})\right]
	\!=\! \frac{\mathrm{tr}\!\left[
		\hat\Pi(\xi_{t_2})e^{i(t_1-t_2)\hat H_E}\hat\Pi(\xi_{t_1})\hat\rho_E(t_1)\hat\Pi(\xi_{t_1})e^{i(t_2-t_1)\hat H_E}
	\right]}{q_E^{(1)}(\xi_{t_1}\xi_{t_1}t_1)}
\!=\! \frac{q^{(2)}_E(\xi_{t_{2}}\xi_{t_{2}}t_{2};\xi_{t_{1}}\xi_{t_1}t_1)}{q_E^{(1)}(\xi_{t_1}\xi_{t_1} t_1)}.
\end{eqnarray}
Therefore, the probability of measuring the sequence of results $(\xi_{t_1},\xi_{t_2})$ (i.e., $\xi_{t_1}$ at $t_1$ followed by $\xi_{t_{2}}$ at $t_{2}$) reads
\begin{eqnarray}
P[(\xi_{t_1},\xi_{t_{2}})] &=& P(\xi_{t_{2}}|\xi_{t_1})P(\xi_{t_1}) = q_E^{(2)}(\xi_{t_{2}}\xi_{t_{2}}t_{2};\xi_{t_{1}}\xi_{t_{1}}t_1),
\end{eqnarray}
and the \textit{a posteriori} state after the latest measurement is
\begin{eqnarray}
\nonumber
\hat\rho(t|\xi_{t_{1}},\xi_{t_{2}}) &=& \frac{1}{P(\xi_{t_2}|\xi_{t_1})}
	e^{-i(t-t_{2})\hat H_E}\hat\Pi(\xi_{t_2})\hat\rho(t_2|\xi_{t_1})\hat \Pi(\xi_{t_2})e^{i(t-t_{2})\hat H_E}\\
&=&\frac{
		e^{-i(t-t_2)\hat H_E}\hat\Pi(\xi_{t_2})
		e^{i(t_1-t_2)\hat H_E}\hat\Pi(\xi_{t_1})\hat\rho_E(t_1)\hat\Pi(\xi_{t_1})e^{i(t_2-t_1)\hat H_E}
		\hat \Pi(\xi_{t_2})e^{i(t-t_2)\hat H_E}
	}{q_E^{(2)}(\xi_{t_2}\xi_{t_2}t_2;\xi_{t_1}\xi_{t_1}t_1)}.
\end{eqnarray}
\end{widetext}
At this point it is clear that when this measurement procedure is iterated over consecutive time steps $t_k>\cdots>t_2>t_1$, the probability to obtain the sequence of results $(\xi_{t_1},\xi_{t_2},\ldots,\xi_{t_k})$ will be given by
\begin{equation}\label{eq:sequence}
P[(\xi_{t_1},\ldots,\xi_{t_k})] = q_E^{(k)}(\xi_{t_k}\xi_{t_k}t_k;\ldots;\xi_{t_1}\xi_{t_1}t_1).
\end{equation}
That is, the probability of measuring the sequence of length $k$ equals the ``diagonal'' part of the quasi-probability $q_E^{(k)}$. Even though, from a technical point of view, the result~\eqref{eq:sequence} was straightforward to arrive at, it is by no means trivial. We have found here a direct link between the dynamics of a system open to $E$ and the process of sequential measurement performed on $E$, without any involvement of open systems. This kind of connection, where the consequences of quantum measurements are quantified with elements of formalism normally employed in open-system theory, has been made previously, albeit in a remote context. The quantum trajectory theory~\cite{Wiseman09} in question, describes the state of a (closed) system that is being \textit{continuously monitored}, i.e., the system that undergoes the sequence of measurements performed with frequency approaching infinity---or, equivalently, infinitesimal delay between consecutive measurements. It is shown there that when the measurement results are ignored by the observer, it is possible to parametrize the time evolution of the \textit{a~posteriori} state of the monitored system with a dynamical equation that is formally identical to the Lindblad form of the master equation that is regularly used to describe open-system dynamics.

The connection with surrogate field trajectories is established when we invoke the noise representation validity~\eqref{eq:validity},
\begin{equation}
P[(\xi_{t_1},\ldots,\xi_{t_k})] \approx p_E^{(k)}(\xi_{t_k}t_k;\ldots;\xi_{t_1}t_1),
\end{equation}
so that probability distribution of the result sequence $(\xi_{t_1},\ldots,\xi_{t_k})$ is the same as the distribution of surrogate field trajectory $\xi(t)$ (a continuous function of time) spanned on a discrete time grid, such that $\xi_{t_1} = \xi(t_1), \xi_{t_2} = \xi(t_2),\ldots, \xi_{t_k} = \xi(t_k)$.

The density of the time grid is, of course, determined by the frequency at which the projective measurements are performed. Therefore, to sample the whole continuity of the trajectories' course, the measurement frequency has to be increased to infinity, in effect, realizing a scheme for continuous monitoring of $\hat V_E$. However, if the surrogate trajectories are being collected for the purposes of simulating the dynamics of open systems, sampling on a discrete time grid is sufficient. On the one hand, if the time grid is fine enough, then the functional integral over continuous trajectories can be approximated by the ensemble average over discrete samples,
\begin{widetext} 
\begin{eqnarray}
\nonumber
\hat\rho_S(t) &=&\int\mathcal{D}\xi\,\mathcal{P}[\xi]\hat U_t[\xi]\hat\rho_S\hat U_t^\dagger[\xi]
\approx \int\mathcal{D}\xi\,\mathcal{P}[\xi]
	\left(\prod_{l=k}^{1} e^{-i(t_{l+1}-t_{l})(\hat H_S+\xi(t_l)\hat V_S)}\right)
		\hat\rho_S
	\left(\prod_{l'={1}}^{k} e^{i(t_{l'+1}-t_{l'})(\hat H_S+\xi(t_{l'})\hat V_S)}\right)\\
\nonumber
&=&\sum_{\xi_k,\ldots,\xi_1\in\Omega}p_E^{(k)}(\xi_k t_k;\ldots;\xi_1 t_1)
	\left(\prod_{l=k}^{1} e^{-i(t_{l+1}-t_{l})(\hat H_S+\xi_{l}\hat V_S)}\right)
		\hat\rho_S
	\left(\prod_{l'=1}^{k} e^{i(t_{l'+1}-t_{l'})(\hat H_S+\xi_{l'}\hat V_S)}\right)\\
&\approx& \frac{1}{N_e}
	\sum_{(\xi_{t_1},\ldots,\xi_{t_k})\in\mathcal{E}_{N_e}}
	\left(\prod_{l=k}^{1} \exp\left[-i(t_{l+1}-t_{l})(\hat H_S+\xi_{t_l}\hat V_S)\right]\right)
		\hat\rho_S
	\left(\prod_{l'=1}^{k} \exp\left[i(t_{l'+1}-t_{l'})(\hat H_S+\xi_{t_{l'}}\hat V_S)\right]\right),
\end{eqnarray}
where $t_{k+1}=t$ and $\prod_{l=k}^1\hat A_l$ ($\prod_{l'=1}^{k}\hat A_{l'}$) is understood as an ordered product $\hat A_{k}\cdots\hat A_1$ ($\hat A_1\cdots\hat A_{k}$).
\end{widetext}
 On the other hand, given the ensemble of sample trajectories, the noise simulation is carried out by first solving the appropriate von Neumann equation for every member of the ensemble with the surrogate trajectories incorporated as an external time-dependent field, and then calculating the ensemble average over so-obtained set of density matrices. To solve the equations of motion in practice, one would employ some numerical method (e.g., the Runge-Kutta method) and, of course, any such method works in discrete time steps---in other words, the numerical solution only requires trajectories to be specified on a time grid. In such a case, the frequency of the measurements---and thus, the density of the grid---would simply set the upper bound for the accuracy of the numerical solutions.

We stress that the ensemble of trajectories sampled with the measurement scheme described above is fit for purpose only when the noise representation is valid [i.e., when~\eqref{eq:validity} and~\eqref{eq:validity_func} are true]. Of course, it is always possible (at least in principle) to perform the sequential measurement to gather results $(\xi_{t_1},\ldots,\xi_{t_k})$, even when $\mathcal{Q}[\xi,\zeta]\neq \delta(\zeta-\xi)\mathcal{P}[\xi]$. The issue is that, unless the noise representation is valid, there is no equivalence between the average over the ensemble of samples and the functional ``average'' over \textit{quasi}-probability distribution $\mathcal{Q}$; such equivalence exists only for proper probability distributions. In other words, measuring the sample trajectories for the explicit purpose of simulating the dynamics of systems coupled to $E$ through $\hat V_E$ can work only when such a simulation can be performed in the first place (i.e., when the noise representation is valid). However, even when~\eqref{eq:validity} does not hold (or it is not clear whether noise representation is valid) but one insists on treating the measured sequences as sample trajectories, there might still be some utility in employing them in a feigned simulation. First, such a simulation could be viewed as a type of approximation to the dynamics of an open system. Second, the feigned simulation could be used as a form of noise representation witness: if one is able to detect that a simulation failed to predict some measurable aspects of the dynamics, then it would constitute a proof that noise representation is not valid. The potential value in the listed use cases might warrant future investigation but, presently, their analysis lies beyond the scope of this work.

\section{Illustration of trajectory sampling in numerical experiment}
Here, we use an example of a numerical experiment to showcase how the noise trajectory sampling and subsequent simulation of open-system dynamics could be implemented in practice.

First, we perform the numerical simulation of sequential measurements of the observable that can be ``replaced'' by the surrogate field; this step is meant to represent the trajectory sampling stage. We take the model of the environment $E$ that can be divided into two subspaces: the two-level system $q$, and the large thermal bath $B$,
\begin{equation}
\hat H_E = \hat H_q\otimes\hat{\mathds{1}}_B  + \hat V_{qB} + \hat{\mathds{1}}_q\otimes\hat H_B.
\end{equation}
We chose to measure the observable that operates only on the subspace $q$,
\begin{eqnarray}
\nonumber
\hat V_E &=& \frac{\omega}{2}\hat\sigma_z \otimes \hat{\mathds{1}}_B = \left(\sum_{\xi=\pm \omega/2} \xi |\mathrm{sign}(\xi)\rangle\langle\mathrm{sign}(\xi)|\right)\otimes\hat{\mathds{1}}_B\\
	&=& \left(\sum_{\xi=\pm \omega/2} \xi\, \hat\Pi(\xi)\right)\otimes\hat{\mathds{1}}_B,
\end{eqnarray}
where $\hat\sigma_z = (|{+}\rangle\langle{+}|-|{-}\rangle\langle{-}|)/2$. Therefore, the spectrum of $\hat V_E$ is $\Omega = \{ +\omega/2, {-}\omega/2\}$ and the corresponding eigenstates are $\{|{+}\rangle, |{-}\rangle\}$. We assume that the relationship between $q$ and $B$ is such that the dynamics of subsystem $q$, $\hat\rho_q(t) = \mathrm{tr}_B(e^{-it\hat H_E}\hat\rho_q\otimes\hat\rho_B e^{it\hat H_E})$, is accurately described by the master equation
\begin{equation}
\frac{d}{dt}\hat\rho_q(t) = \mathscr{L}\hat\rho_q(t) = -\frac{\gamma}{2}[\hat\sigma_x,[\hat\sigma_x,\hat\rho_q(t)]],
\end{equation}
so that $\hat\rho_q(t) = e^{(t-s)\mathscr{L}}\hat\rho_q(s)$ for any $t>s>0$. For simplicity we also assume that the state of $q$ is stationary, $\mathscr{L}\hat\rho_q = 0$, which implies that $\hat\rho_q = \hat{\mathds{1}}/2$. It has been demonstrated in~\cite{Szankowski20} that such an environment facilitates valid noise representation where the observable $\hat V_E$ is replaced by the surrogate field $\xi$ in a form of \textit{random telegraph noise}---a stochastic process that switches between $\pm\omega/2$ at rate $\gamma$. 

For this model, the probability of getting result $\xi_{t_1}$ in the measurement of $\hat V_E$ simplifies as follows
\begin{eqnarray}
\nonumber
P(\xi_{t_1}) &=& \mathrm{tr}\left[\hat\Pi(\xi_{t_1})\otimes\hat{\mathds{1}}_B\, e^{-it_1\hat H_E}\hat\rho_q\otimes\hat\rho_B\,e^{it_1\hat H_E}\right]\\
&=&\mathrm{tr}_q\left[\hat\Pi(\xi_{t_1})e^{t_1\mathscr{L}}\hat\rho_q\right],
\end{eqnarray}
so that $B$ degrees of freedom are eliminated from the problem. Therefore, it is enough to consider the \textit{a posteriori} state reduced to subspace $q$, 
\begin{equation}
\hat\rho_q(t|\xi_{t_1}) = e^{(t-t_1)\mathscr{L}}\frac{\hat\Pi(\xi_{t_1})\hat\rho_q(t_1)\hat\Pi(\xi_{t_1})}{P(\xi_{t_1})}.
\end{equation}
In general, the \textit{a posteriori} state after $l$ measurements is given by the recurrence relation,
\begin{eqnarray}
\nonumber
&&\hat\rho_q(t | \xi_{t_1},\ldots,\xi_{t_l})\\
\label{eq:num_a_posteriori}
&&\phantom{\rho_q}
 = e^{(t-t_l)\mathscr{L}}\frac{
		\hat\Pi(\xi_{t_l})\hat\rho_q(t_l|\xi_{t_1},\ldots,\xi_{t_{l-1}})\hat\Pi(\xi_{t_l})
	}{
		P(\xi_{t_l}|\xi_{t_1},\ldots,\xi_{t_{l-1}})
	},
\end{eqnarray}
with the conditional probability of getting the result $\xi_{t_l}$ that reads
\begin{eqnarray}
\nonumber
P(\xi_{t_l}|\xi_{t_1},\ldots,\xi_{t_{l-1}}) = \mathrm{tr}_q[\hat\Pi(\xi_{t_l})\hat\rho_q(t_l|\xi_{t_1},\ldots,\xi_{t_{l-1}})].\\[-.05cm]
\label{eq:num_prob}
\end{eqnarray}

\begin{figure}[t]
\centering
\includegraphics[width=\columnwidth]{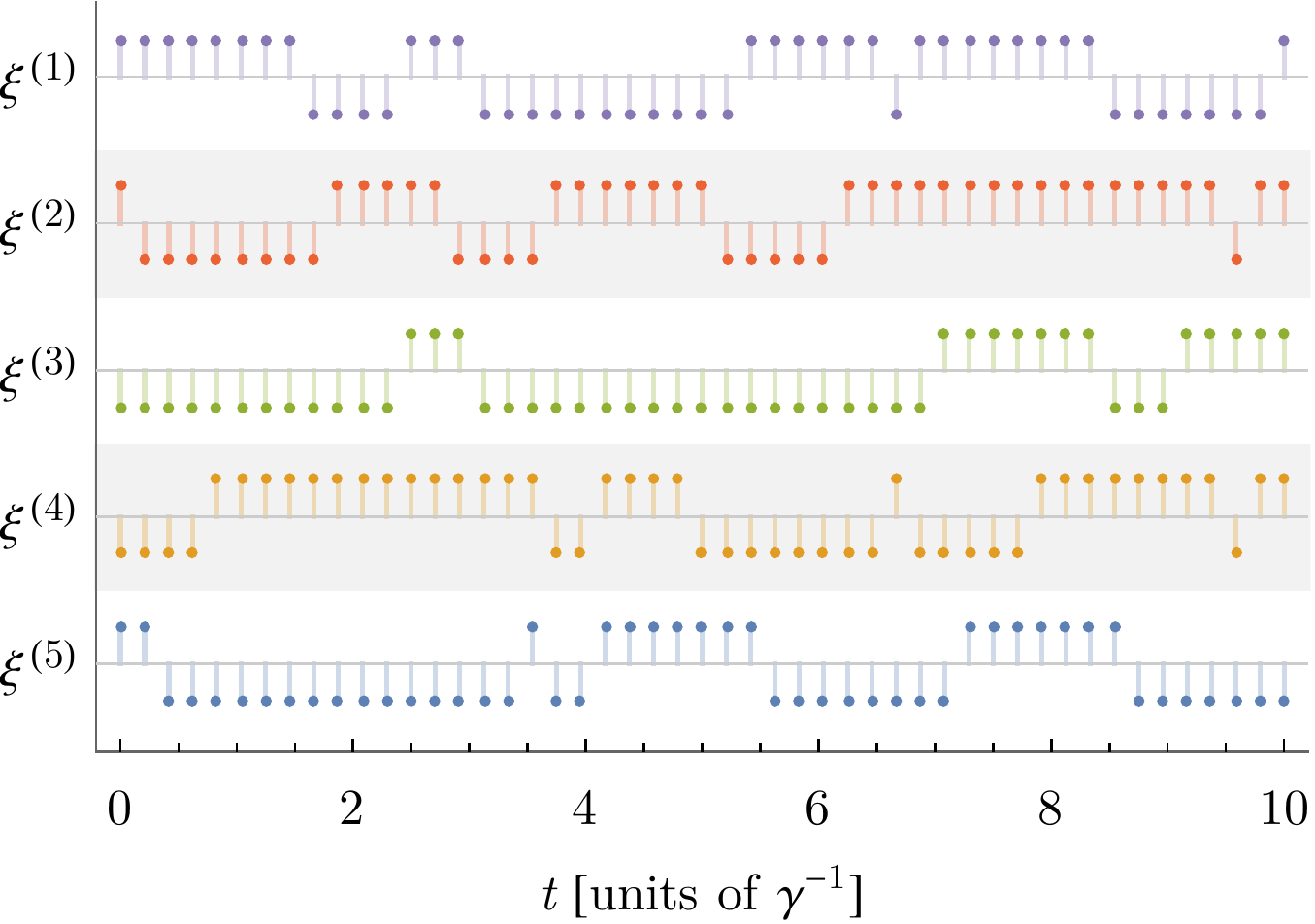}
\caption{First five sample trajectories $\xi^{(j)}$ of the surrogate field $\xi$ measured and stored in the ensemble $\mathcal{E}_{N_e}$ in the numerical experiment. The trajectories switch at random between two values $\pm \omega/2$. In this example, the projective measurements were carried out every $0.2\,\gamma^{-1}$.}
\label{fig:samples}
\end{figure}

The numerical simulation of the sequential measurement is implemented in the following way. We choose the interval between consecutive measurements $\Delta = 0.2\times\gamma^{-1}$ and the noise strength $\omega=2\gamma$ ($\gamma^{-1}$ is used here as the unit of time). Then, we start by drawing the first value $\xi_{t_1}^{(1)}$ at $t_1=0\times \Delta$ from calculated probability distribution $P(\xi_{t_1}^{(1)})$ (recall, the initial state of $E$ is stationary). Once the value of the first result is set, we then calculate the \textit{a posteriori} state $\hat\rho_q(\Delta|\xi^{(1)}_{t_1})$ according to Eq.~\eqref{eq:num_a_posteriori}, which, in this case, can be done analytically. Using this state we calculate the probability distribution $p(\xi^{(1)}_{t_2}) = P(\xi^{(1)}_{t_2}|\xi_{t_1}^{(1)})$ according to formula~\eqref{eq:num_prob} (again, can be done exactly) and we draw from it the next result $\xi_{t_2}$ at $t_2=\Delta$. These steps are repeated $48$ more times (for the total of $k=50$ measurements) and we obtain the sequence $(\xi^{(1)}_{t_1},\xi^{(1)}_{t_2},\ldots,\xi^{(1)}_{t_{50}})$---the first sample trajectory. The whole procedure is repeated $1000$ times and each obtained sample is stored in the ensemble $\mathcal{E}_{1000} = \{ (\xi_{t_1}^{(j)},\ldots,\xi_{t_{50}}^{(j)})\}_{j=1}^{1000}$; the obtained samples are plotted in Fig.~\ref{fig:samples}.

Now that we have populated the ensemble of trajectory samples for our model environment, we can make use of it to calculate the noise representation simulation of the open-system dynamics of any $S$ coupled to $E$ via $\hat V_E$. For this demonstration, we have chosen $S$ as a qubit interacting with the environment via so-called pure dephasing coupling,
\begin{equation}
\hat H_S = 0;\ \hat V_S = \frac{1}{2}\hat\sigma_z;\ \hat\rho_S = \frac{1}{2}\left(\begin{array}{cc}1&1\\1&1\\\end{array}\right),
\end{equation}
so that
\begin{equation}
\hat H_{SE} = \frac{1}{2}\hat\sigma_z \otimes\hat V_E + \hat{\mathds{1}}\otimes\hat H_E \to \hat H_t[\xi] = \frac{1}{2}\xi(t)\hat\sigma_z.
\end{equation}
Pure dephasing implies that only the off-diagonal elements of the density matrix are evolving, $\langle {\pm}|\hat\rho_S(t)|{\pm}\rangle = \mathrm{constant}$. The choice of $S$ was dictated by the fact that the dynamics can be solved exactly and the state of the system is given by the analytical formula~\cite{Ramon12,Szankowski13}, which can be easily compared with the ensemble average.

\begin{figure}[t]
\centering
\includegraphics[width=\columnwidth]{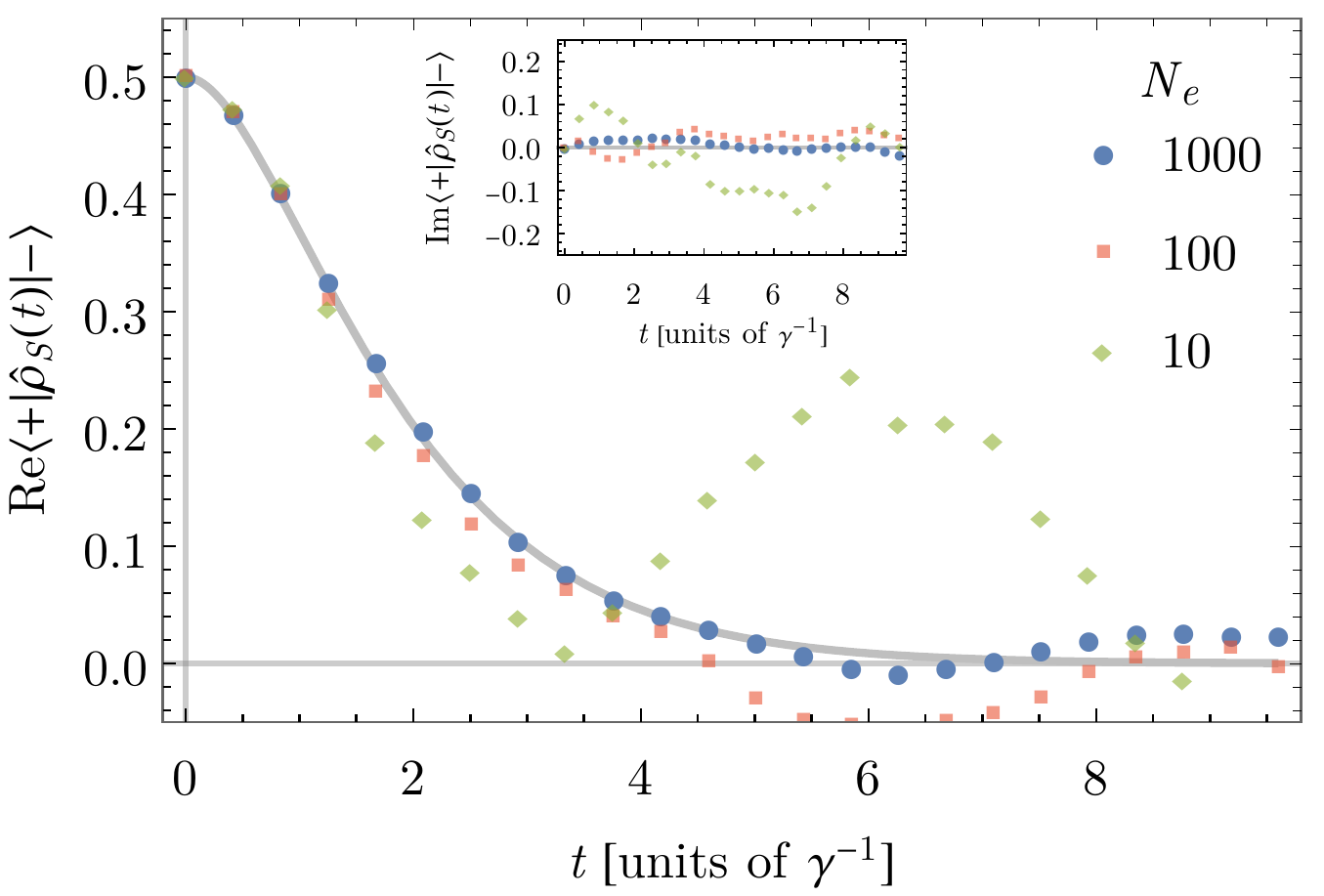}
\caption{The simulated evolution of the system's coherence $\langle{+}|\hat\rho_S(t)|{-}\rangle$. The plot compares the exact result $(1/2)e^{-\gamma t}\left[\cosh(\mu \gamma t)+\mu^{-1}\sinh(\mu \gamma t)\right]$ with $\mu = \sqrt{1-\omega^2/(4\gamma^2)}$ (solid line) and ensemble averages $\hat\rho_S(t)\approx (1/N_e)\sum_{j=1}^{N_e}\hat\rho_S^{(j)}(t)$ obtained for three values of ensemble size $N_e$: $10$ (green diamonds), $100$ (red squares), and $1000$ (blue circles). The trajectory-wise density matrices $\hat\rho_S^{(j)}(t)$ were obtained by numerical integration of corresponding von Neumann equation~\eqref{eq:example_vonNeumann}.}
\label{fig:averages}
\end{figure}

We employ the trajectory ensemble to perform the simulation by solving the von Neumann equation for each measured sample,
\begin{eqnarray}
\nonumber
&&\hat\rho_S^{(j)}(0) = \frac{1}{2}\left(\begin{array}{cc}1&1\\1&1\\\end{array}\right),\\
\label{eq:example_vonNeumann}
&&\frac{d}{dt}\hat\rho_S^{(j)}(t) = -\frac{i}{2}\xi^{(j)}(t) [ \hat\sigma_z , \hat\rho_S^{(j)}(t) ],
\end{eqnarray}
using Runge-Kutta method with time step $h=2\Delta$ and substituting the trajectory values with the corresponding measured results, $\xi^{(j)}(l \Delta) = \xi_{t_l}^{(j)}$. (Note that the Runge-Kutta algorithm requires trajectories evaluated at $t=n h = 2 n \Delta$ and $t = n h +h/2 = (2n+1)\Delta$, hence the trajectories where sampled every $\Delta = h/2$.) Obtained density matrices are collected in the ensemble $\{\hat\rho_S^{(j)}(t)\}_{j=1}^{1000}$; the results of averages calculated with first $10$, $100$ and full $1000$ elements are shown in Fig.~\ref{fig:averages}. The accuracy of the noise representation simulation improves with the size of the ensemble. Overall, the ensemble average exhibits fluctuations around the exact value with rms that scales as $1/\sqrt{N_e}$---the shot noise scaling due to averaging with a finite number of sample trajectories. However, the error is not uniform in $t$: at short times scales ($t\sim\gamma^{-1}$, $\omega^2 t\lesssim \gamma$) the error is small even for modest ensemble sizes, and it grows to maximal value at long times that coincide with the significant decay of the coherence $\langle{+}|\hat\rho_S(t)|{-}\rangle\sim 0$.

\section{Trajectory sampling and noise spectroscopy: a comparison}\label{sec:vs_spectro}

In essence, the trajectory sampling is a way to characterize the surrogate field through direct measurement---a scheme for ``measuring'' the environmental noise. However, the concept of ``measuring'' the noise is not new; the most popular technique of this sort is the dynamical-decoupling-based noise spectroscopy~\cite{Paz16, Szankowski17, Paz17, Ferrie18, Szankowski18, Norris18, Krzywda19, Szankowski19}. It is instructive to compare and contrast the two approaches. To this end, we first give a brief rundown of the principle of operation for the spectroscopy.

Noise spectroscopy performs its ``measurement'' indirectly by utilizing a simple qubit probe ($p$) brought in contact with the environment through pure dephasing,
\begin{eqnarray}
\nonumber
    &&\hat H_{pE}(t) = \frac{1}{2}f_\mathrm{ctr}(t)\hat\sigma_z\otimes\hat V_E+\hat{\mathds{1}}\otimes\hat H_E \\
    &&\phantom{H}\to \hat H_t[\xi] = \frac{1}{2}f_\mathrm{ctr}(t)\xi(t)\hat\sigma_z,
\end{eqnarray}
where we assumed that $E$ has a valid noise representation with surrogate field $\xi$. The real-valued filter function $f_\mathrm{ctr}(t)$ encapsulates the effects of the control protocol exerted over the qubit. Typically, the employed control scheme has the form of a sequence of precisely timed pulses that causes effectively instantaneous $\pi$ rotations of the qubit's Bloch vector (spin flips)---an approach inspired by the dynamical decoupling techniques. The control is the key element of the method; usually, the control is designed to exert a periodic modulation with a well-defined frequency $\omega_\mathrm{ctr}$, so that $f_\mathrm{ctr}(t)$ can act as narrow pass-band frequency filter. The pure dephasing coupling is chosen because it leads to very simple qubit evolution where the filter action is easily manipulated and the rate of qubit's dephasing (the decay of off-diagonal elements of the density matrix) is most directly related to noise-characterizing quantities,
\begin{eqnarray}
\nonumber
    \langle{+}|\hat\rho_p(t)|{-}\rangle &\propto&
    \exp\left[
        \sum_{k=1}^\infty \frac{(-i)^k}{k!}\int_0^tdt_1\cdots dt_k\right.\\
\label{eq:spectro_coherence}
&&\left.\times\left(\prod_{l=1}^k f_\mathrm{ctr}(t_l)\right)
        \overline{\overline{\xi(t_1)\cdots\xi(t_k)}}
    \right].
\end{eqnarray}
Here, $\overline{\overline{\xi(t_1)\cdots\xi(t_k)}}$ is the cumulant of order $k$ of stochastic process $\xi$~\cite{VanKampen11}. Formally, the set of cumulants of all orders (in general, there are infinitely many orders) fully defines the stochastic process. The cumulant of a given order $k$ can always be expressed as a combination of moments $\overline{\xi(t_1)\cdots\xi(t_l)}$ [see Eq.~\eqref{eq:moment}] of orders $l\leqslant k$, even though the explicit formulas get progressively more complex; e.g., the second cumulant, called the autocorrelation function, is given by
\begin{equation}
    C(t_1,t_2) = \overline{\overline{\xi(t_1)\xi(t_2)}}=\overline{\xi(t_1)\xi(t_2)} - \overline{\xi(t_1)}\,\overline{\xi(t_2)}.
\end{equation}
The stated objective of noise spectroscopy is to characterize the noise by recovering its cumulants. We illustrate how this approach works in practice with a simple example: Consider the case when the qubit's response~\eqref{eq:spectro_coherence} is well described within Gaussian approximation where the contributions from cumulants of order $>2$ are negligible. In addition, assume the noise is a stationary process so that $C(t_1,t_2) = C(t_1-t_2)$. When the filter frequency $\omega_\mathrm{ctr}$ of the applied control sequence satisfies $\omega_\mathrm{ctr}\gg 2\pi \tau_c^{-1}$ where $\tau_c$ is the noise correlation time [the range of $C(t)$] and the total duration is set to $t = 2\pi n\omega_\mathrm{ctr}^{-1}$ where the integer $n\gg 1$ so that $t\gg \tau_c$, then the dephasing rate is given by~\cite{Szankowski18}
\begin{equation}
    \langle{+}|\hat\rho_p(t)|{-}\rangle \approx \langle {+}|\hat\rho_p(0)|{-}\rangle\exp\left[-\frac{4t}{\pi^2}S\left(\omega_\mathrm{ctr}\right)\right],
\end{equation}
where $S(\omega) = \int_{-\infty}^\infty e^{-i\omega s}C(s)ds$ is the noise power spectral density---the Fourier transform of the autocorrelation function. Therefore, by measuring the decay rate for a wide range of filter frequencies $\omega_\mathrm{ctr}$ one performs a tomographic reconstruction of spectral density (hence, the name noise \textit{spectroscopy}). Since $S(\omega)$ contains essentially the same information as $C(t)$, the final step of reverting back to autocorrelation function through inverse Fourier transform is usually skipped at this point.

Typically, the noise spectroscopy is utilized to extract information that fills the gaps in the assumed model of the environment (e.g., positions of spectral lines) \cite{Lovchinsky16, Krzywda17} and only rarely can it be used to characterize the surrogate field with enough detail that would allow for the obtained information to be reused for the purpose of simulating the dynamics of other open systems. The reason is that, even though the full set of cumulants defines the process completely, converting cumulants into an ensemble of trajectories is an excessively difficult process (e.g., using the multivariate version of Gram--Charlier A series) that is fraught with significant problems with convergence and accuracy. Naturally, the quality of such a conversion procedure is directly tied to the number of available cumulants; this turns out to be the major drawback because, in practice, the noise spectroscopy methods can feasibly reconstruct only a few first cumulants~\cite{Norris16, Ramon19}. In fact, the standard designs of spectroscopy protocols do not work without Gaussian approximation.

It is clear that a method for populating the trajectory ensemble through direct measurement solves the issue of the conversion procedure by simply circumventing it entirely. Moreover, the ensemble assembled in such a way not only enables the simulation of open-system dynamics, but it also allows us to calculate any quantity characterizing the noise itself, including moments, power spectral density, and cumulants of arbitrary order (Gaussian approximation is unnecessary); e.g., the autocorrelation function can be obtained with the following formula:
\begin{eqnarray}
\nonumber
C(t_1,t_2) &\approx& \frac{1}{N_e}\sum_{j=1}^{N_e}\xi^{(j)}(t_1)\xi^{(j)}(t_2)\\
\label{eq:ensemble_cum}
&& - \frac{1}{N_e^2}\sum_{j,j'=1}^{N_e}\xi^{(j)}(t_1)\xi^{(j')}(t_2),
\end{eqnarray}
where $\xi^{(j)}\in\mathcal{E}_{N_e}$. Therefore, the trajectory sampling method could be considered as a next evolutionary step of noise spectroscopy.

\section{Conclusions}

We have shown that, for the environment $E$ that supports the noise representation $\xi$ of observable $\hat V_E$ [i.e., when, for an arbitrary system $S$ coupled to $E$ via $\hat V_E$, the dynamics of $\hat\rho_S(t)$ can be simulated with stochastic one-body Hamiltonian $\hat H_t[\xi]$, see Eq.~\eqref{eq:H_SE}], the result of sequential projective measurement of $\hat V_E$ has a probability distribution identical to trajectories of surrogate field~$\xi$. Therefore, even though it is impossible to observe fluctuation of the noise field with a probe coupled to $E$ [if $S$ is a probe, then $\hat\rho_S(t)$ is averaged over fluctuations of $\xi$], we have demonstrated that, effectively, a single noise trajectory can be exposed to direct observation through sequential measurement performed on $E$.

If noise trajectories can be measured, then they can be sampled and stored for later use. One use case is to characterize the noise itself with quantities such as moments, power spectral densities, or cumulants, all of which can be easily calculated with sample trajectories [e.g., see Eq.~\eqref{eq:ensemble_cum}]. Thus, trajectory sampling can be viewed as an upgrade to standard noise spectroscopy that circumvents typical limitations, most notably, the reliance on Gaussian approximation. 

Here, we have focused on the utility of trajectory sampling in simulating the dynamics of systems open to $E$. When the noise representation is valid, the reduced state of open system is found by averaging the conditional unitary evolution over surrogate field trajectories [see, Eq.~\eqref{eq:path_int}]. However, computing the necessary probability distribution from first principles is prohibitively difficult and requires the full knowledge of $E$. Moreover, even when the distribution is supplied, averaging over it is technically impractical. All these difficulties are circumvented with trajectory sampling: (i) the average over distribution can be approximated by the average over finite number of samples (the ensemble average); (ii) the samples are obtained from direct measurement, and thus, the knowledge about environmental dynamics is not necessary.

There are two main reasons why trajectory sampling was not considered before. First, so far, there was no compelling argument against the view that the trajectories are nothing more than artifacts of the noise representation, and as such, they would be impossible to observe or measure. Indeed, the observable quantities obtained through surrogate field simulation always involve the average over trajectories [see Eq.~\eqref{eq:rhoS}]. Therefore, there is no measurement that could be performed on the system state $\hat \rho_S(t)$ that would be able to somehow skip the averaging and expose a single trajectory of the noise. Moreover, given that the noise is, by its very definition, only a surrogate for actual quantum system $E$, there seems to be no reason to think of $\xi$ as anything more than a mathematical construct used in an intermediate step of calculations that only at the terminus give physically meaningful result. The second reason is that even if trajectories could be measured somehow, it was unclear whether the noise that generated them could be repurposed to simulate the dynamics of systems other than the one used to carry out the measurement in the first place (i.e., the issue of surrogate field's objectivity). Presently, we know that the surrogate field is an objective representation~\cite{Szankowski20}, and, thus, the latter objection can be put to rest. The motivation for assembling the trajectory ensemble is sound, as the ensemble can be utilized for setting up simulations of any system exposed to the environment through coupling $\hat V_E$. The former issue, however, is more subtle. Here, we have not settle the question whether noise trajectories are ``real'' or ``artificial.'' In fact, we have not even engaged with this question. What we have shown is that the result of a properly set up sequential measurement of observable $\hat V_E$ has a probability distribution identical to that of the hypothetical noise trajectory. On the one hand, the fact that trajectories are identically distributed as the measured sequences allows one to substitute the latter for the former (and \textit{vice versa}) when performing the ensemble average. On the other hand, the equality of probability distributions does not imply that the noise trajectory and the measured sequence are the same entity, and thus, the trajectory cannot simply ``inherit'' tangibility from the measured sequence. After all, physical setups required to perform the sequence of measurements differ substantially from the setup where the object of investigation is the system state $\hat\rho_S(t)$---in particular, the sequential measurement does not involve system $S$.

\section*{Acknowledgments}
I would like to thank J.~Krzywda and \L{}.~Cywi\'{n}ski for many insightful discussions and for reviewing drafts of this paper.

\end{document}